\documentstyle[psfig,aps,twocolumn]{revtex}
\newcommand{\s}{\mathrm}

\newcommand{\ra}{\rightarrow}
\newcommand{\mn}{\mu \nu}
\newcommand{\be}{\begin{equation}}
\newcommand{\ee}{\end{equation}}

\newcommand{\ba}{\begin{eqnarray}}
\newcommand{\ea}{\end{eqnarray}}
\newcommand{\bef}{\begin{figure}}
\newcommand{\eef}{\end{figure}}

\newcommand{\ep}{\epsilon}
\newcommand{\lda}{\lambda}

\newcommand{\opr}{\omega \pi \rho}
\tightenlines
\begin{document}
\draft
\title{Omega meson as a chronometer and thermometer in hot-dense
hadronic matter.}
\author{Pradip Roy$^1$, Sourav Sarkar$^1$, Jan-e Alam$^1$
Binayak Dutta-Roy$^2$ and Bikash Sinha$^{1,2}$}
\address{$^1$Variable Energy Cyclotron Centre,
     1/AF Bidhan Nagar, Calcutta 700 064
     India}
\address{$^2$Saha Institute of Nuclear Physics,
           1/AF Bidhan Nagar, Calcutta 700 064
           India}

\maketitle

\begin{abstract}
Changes in the properties of the vector mesons in hot and dense 
hadronic matter, as produced in heavy ion collisions, lead to 
the intriguing possibility of the opening of the decay channel 
$\omega\,\ra\,\rho\,\pi$, for the omega meson, which is impossible 
in free space. This  along with the channel $\omega\,\pi\,\ra\,\pi\,\pi$  
would result in a decrease in its effective life-time enabling it to 
decay within the hot zone and act as a chronometer in contradiction 
to the commonly held opinion and would have implications vis a vis 
determination of the size of the region through pion interferometry.
A new peak and a radically altered shape 
of the low invariant mass dilepton 
spectra appears due to different shift in the masses of $\rho$ and 
$\omega$ mesons.
The Walecka model is used for the underlying calculation for the sake of 
illustration.
\end{abstract}
\vskip 0.2in

It is well established by now that the essential properties of the hadrons,
once immersed in hot dense hadronic matter do change perceptibly.
The spectral analysis of particles ejected from relativistic heavy
ion collisions can give some insight into the physical
properties of the system; any change in the spectrum 
offers an exciting possibility to study the structure of hadrons
and the QCD ground state at finite temperature and density. The 
light vector mesons ($\rho$ and $\omega$) can act as indicators for the 
partial restoration of chiral symmetry, a symmetry which is broken
in the hadronic ground state. 
Reliable informations on the evolving state of this hot and dense
strongly interacting matter would hardly be possible from hadronic 
signals as these would be
masked behind layers of complex dynamics. On the other hand, such inferences
could perhaps be drawn from the dilepton ($l^+l^-$) spectrum which does not
suffer from such strong distortions and carries information about the 
environment of the hadrons from which the pairs originate. These
electromagnetic probes couple to hadrons through spin one ($J^{P}=1^-$)
mesons. Final spectra exhibit a resonant structure which, in the low mass
regime includes the rho ($\rho$) and the omega ($\omega$). The isovector
rho meson of mass ($m_\rho$) 770  MeV has a full width ($\Gamma_\rho$) of 151
MeV which is mainly accounted for through the two pion decay channel.
The isoscalar $\omega$ of mass ($m_\omega$) 782 MeV has a far narrower
width ($\Gamma_\omega$) of 8.4 MeV since the two pion mode is forbidden
by G-parity conservation and is 
allowed to decay into three pions with a consequent substantial
reduction in phase space. It is generally believed that because of the
narrow width of omega, which dictates a long lifetime
($\tau_\omega\sim 23 $ fm/c) as compared to
that of the $\rho$ ($\tau_\rho\sim 1.3 $ fm/c),
the former (in contrast to the $\rho$) decays outside the hot and
dense region, and thus while $\rho \ra l^+l^-$
does provide information on the fireball, $\omega \ra l^+l^-$
does not. 

However, it is the contention of this letter,
that this is not necessarily so. 

Changes in the masses and
decay widths in nuclear matter at finite temperature and density may
indeed radically alter the scenario as would be argued below. Thus,  
both $\rho$ and $\omega$ may serve as sensitive chronometers
and thermometers of the evolving hadronic gas.
We wish to draw attention to the fact that qualitatively interesting
and amusing phenomena can occur in any scenario where the masses
of different hadrons behave differently as a function of temperature
and density. 
Thus under suitable conditions of temperature and density
of its environment, the mass of the omega can exceed the sum 
of the masses of the $\rho$ and the $\pi$ and thereby the two particle 
$\rho\pi$ channel can open up. In a hadronic medium the channel 
$\omega\,\pi\,\ra\,\pi\,\pi$ is also very important for the depletion 
of $\omega$. Due to these two processes along with $\omega\,\ra\,3\pi$ 
the narrow $\omega$ can become dramatically broad in the medium. 

Many authors~\cite{Hatsuda,Brown,Pisarski} have investigated the 
issue of temperature dependence of
hadronic masses within different models over the past several years.
We employ one of the most
extensively used and well tested models as far as nuclear 
matter calculations are concerned, namely the Walecka model~\cite{vol16}.
The Walecka model comprises of the scalar $\sigma$,
the $\rho$, $\omega$ and nucleon fields interacting through the
Lagrangian
\ba
{\cal L}&=&g_{\sigma}{\bar N}\phi_{\sigma}N
-g_{V N N}\left[\frac{}{}{\bar N}\gamma_\mu \tau^\alpha N
\right.\nonumber\\ 
&&\left.-i\frac{\kappa_{V}}{2M}{\bar N} \sigma_{\mn}\tau^\alpha N
\partial^\nu\right]V_\alpha^\mu 
\ea
where, $\phi_\sigma$ and $N$ are the sigma and nucleon fields
and the generic vector field is denoted by $V_\alpha^\mu$,
$\alpha$ running from 0 to 3, indexes quantities relevant 
for $\omega$ (when $\alpha = 0$) and for $\rho$ ($\alpha$ = 1 to 3);
also $\tau_0 $ = 1 and $\tau_i$ are the isospin Pauli matrices.
The value of the $\sigma$ mass has been taken to be $m_\sigma$ =
450 MeV and the coupling constants $g_{\omega N N} \sim$ 10,
$g_{\rho NN} \sim$ 2.6, $\kappa_{\rho} \sim$ 6.1, $\kappa_{\omega} =$ 0
and $g_{\sigma} \sim$ 7.4, chosen so as to reproduce the saturation
density and the binding energy per nucleon in nuclear matter.

The effective nucleon mass (which appears in the nucleon loop 
contribution to self energies of the rho and 
omega mesons) has been calculated, within the framework of 
the model defined above, in the Relativistic Hartree Approximation.  
The major contribution to the medium effects on the rho and omega mesons,
in this approach, arises from the nucleon-loop diagram.
For the dressing of internal lines in matter we restrict ourselves to 
Mean Field Theory (MFT) to avoid a plethora of diagrams and to maintain
internal consistency. 
It has been shown earlier~\cite{GS} that the change in rho 
mass due to $\pi-\pi$ loop is negligibly small at nonzero
temperature and zero density. 
It is found that the in-medium mass of $\omega$ ($m_\omega^\ast$)
decreases less rapidly than that of $\rho$ ($m_\rho^\ast$)
with temperature (T) and density ($n_B$) such that in a region 
of $n_B$ and $T$ (to be delimited later) the $\omega\,\ra\,\rho\,\pi$
channel which in free space is kinematically forbidden
because $m_\omega\,<\,m_\rho\,+\,m_\pi$ becomes possible in matter
when $m_\omega^\ast\,>\,m_\rho^\ast\,+\,m_\pi$. Within the framework
of the model adopted here,
the mass of the rho meson decreases more rapidly than the mass of
the omega because 
they couple to nucleons with different coupling strength,
as is evident from the values of the coupling constant
quoted in the previous paragraph. We do not observe any universal
scaling law~\cite{Mrho} in our calculation. Different rate of 
variation for rho and omega masses with temperature 
has been reported in Ref.~\cite{Klingl} also.
Since the Walecka model does not have chiral symmetry,
it is rather difficult to predict something reliable
on the pion mass in this model, especially in the MFT approximation
~\cite{vol16,Kapusta}.
On the other hand, in the models with chiral symmetry
e.g. the Nambu-Jona-Lasinio model, and the linear sigma model
with nucleon, it is well-known that
the pion mass is almost unchanged in so far as one is in the 
Nambu-Goldstone phase. This is simply a consequence of the
Nambu-Goldstone theorem in medium~\cite{Hatsuda1};
we thus adopt this approach of keeping the pion mass constant.
\vskip .1in
\bef
\centerline{\psfig{figure=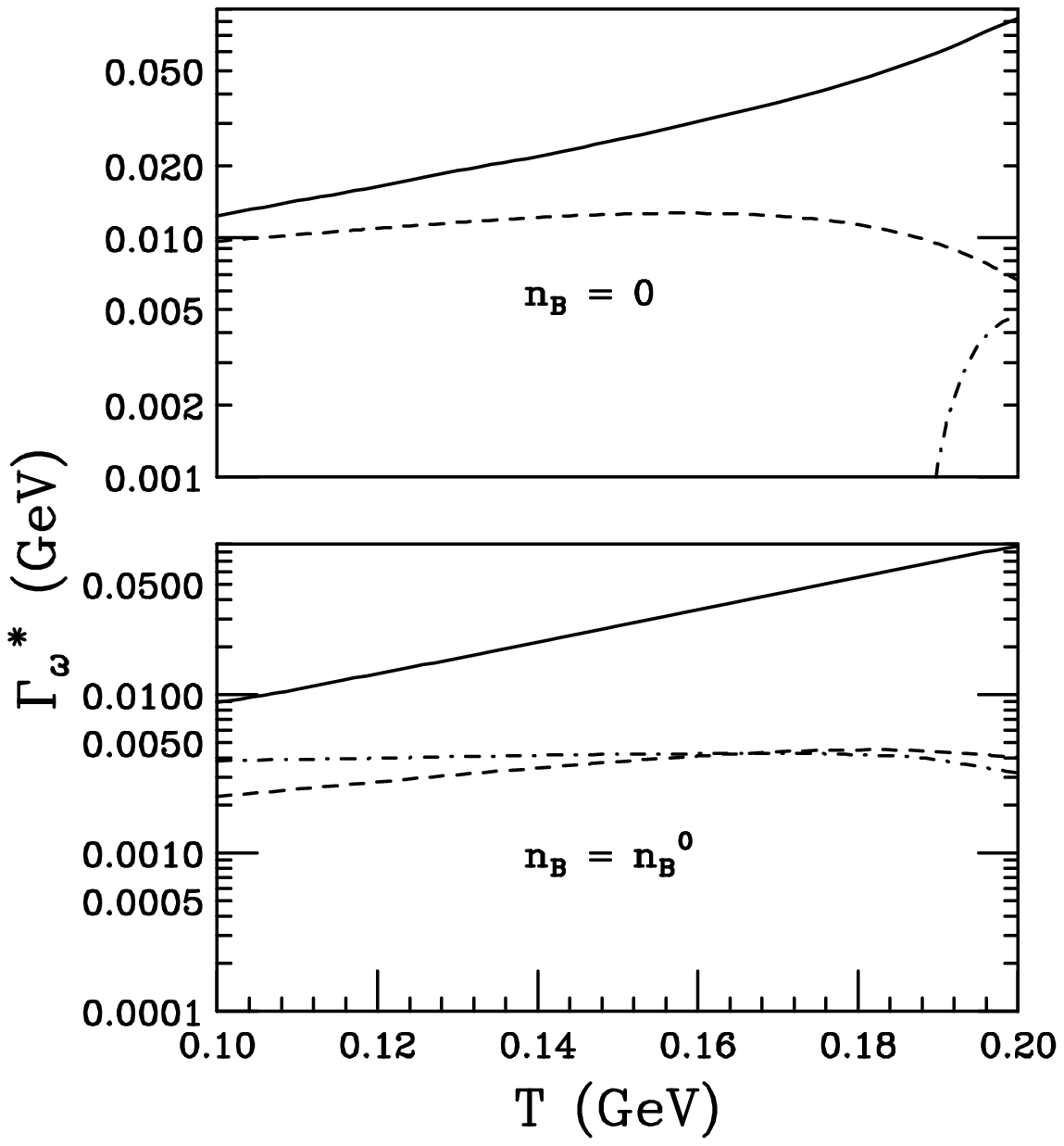,height=10cm,width=8cm}}
\caption{In-medium decay widths of omega meson for 
$\omega\,\ra\, \rho\,\pi$ (dot-dashed line) and
$\omega\,\ra\,3\pi$ (dashed line) as a function of temperature at zero and
normal nuclear density. The solid line
indicates the total in-medium depletion rate of $\omega$ due to
the above two processes along with the contribution from
$\omega\,\pi\,\ra\,\pi\,\pi$.  
}
\label{wcomp}
\eef
\vskip .1in

In order to delimit
the extremities, it may be appropriate to remark that the
decay $\omega\,\ra\,\rho\pi$ is possible for a baryon density
$n_B/n_B^0\geq 0.3$ ($n_B^0$ being the normal nuclear matter density)
at zero temperature, whereas at zero
baryon density this is allowed for $T\geq 185$ MeV. 
The width for the decay $\omega\,\ra\,\rho\,\pi$ when allowed is
calculated by using the Lagrangian proposed by Gell-Mann, Sharp and Wagner
(GSW)~\cite{Gell}
\be
\cal L=\frac{g_{\opr}}{m_\pi}\ep_{\mu \nu \alpha \beta}\partial^\mu
\omega^\nu\partial^\alpha{\vec\rho}^\beta\cdot{\vec\pi}
\ee
and employing for the in-medium decay width the finite temperature 
cutting rules, one has
\ba
\Gamma_{\omega\ra\rho\pi}^{\ast}&=&\frac{g_{\opr}^2}
{32\pi m_{\omega}^{\ast 3}
m_{\pi}^2}\lda^{3/2}(m_{\omega}^{\ast 2},m_{\rho}^{\ast 2},m_{\pi}^2)
\nonumber\\
&&\times\,\left[\frac{}{}1+f_{BE}(E_\pi) 
+f_{BE}(E_\rho)\right]
\ea 
where $\lambda$ is the appropriate phase space factor (triangular
function), while $f_{BE}$ is the Bose-Einstein distribution for the
pions and the rho mesons in equilibrium. 
The coupling constant $g_{\omega\rho\pi}\sim 2$ can be deduced 
from the observed decay $\omega\ra\,\pi^0\,\gamma$ using the
vector dominance model of Sakurai~\cite{VMD} for 
the $\rho\gamma$ vertex taking
the process to occur through a virtual rho which converts to a
photon. The three body decay width for $\omega\,\ra\,3\pi$
is estimated from the phenomenological
effective Lagrangian~\cite{Sakurai}
\be
\cal L_{\omega 3\pi} = 
f_{\omega 3\pi}\ep_{\mn\alpha\beta}\omega^\mu\epsilon^{ijk}
\partial^\nu\pi_i\partial^\alpha\pi_j\partial^\beta\pi_k
\ee
the latin indices referring to isospin, avoiding for the present
purpose the GSW model where this decay
proceeds via a virtual rho ($\omega\,\ra\,\rho\,\pi\,\ra\,\pi\,\pi\,\pi$)
in order to avoid the possibility of double counting when the threshold
for the two body decay is crossed. 

The resulting depletion rate of omega  as a function of temperature at zero 
baryon density ($n_B=0$) and at normal nuclear densities ($n_B=n_B^0$)
is depicted in 
Fig.~\ref{wcomp}; and it may be  noted that the two body channel 
opens up in the 
former case at a temperature $\sim 185$ MeV, while in the latter situation 
(normal nuclear density) this channel remains open even at zero temperature.
The opening of the channel $\omega\,\ra\,\rho\,\pi$ and the process
$\omega\,\pi\,\ra\,\pi\,\pi$ endows the omega-meson with a width
comparable to that of the $\rho$ meson 
as a result of which the lifetime of the omega meson reduces 
to $\sim 2.3$ fm/c which is comparable to the rho meson lifetime 
($\sim 2.1$ fm/c) under the  same condition. Thus
the omega meson gets promoted to be an effective probe for
the early stages of hadronic matter formed in relativistic heavy 
ion collisions. This is in sharp contradiction to the commonly held
notion according to which the omega is too long-lived \cite {Heinz,Shuryak} 
to serve this purpose.
However, as mentioned earlier the hadronic decay modes of rho and
omega in the fire-ball are not very informative and that they 
are experimentally `visible', so to say, through their dileptonic
decay modes. Therefore, it is more relevant to evaluate the dilepton
emission rate from the decay of unstable vector particles ($\rho$ and $\omega$)
by using the generalized  Breit-Wigner formula at a non-zero temperature
and density \cite{Weldon} 
\ba
\frac{dR}{dM}&=&\frac{2J+1}{\pi^2}\,M^2T\,\sum_n\,
\frac{K_1(nM/T)}{n}\nonumber\\
&&\times\,\frac{m_V^{\ast}\,\Gamma_{\s{tot}}^{\ast}
/\pi}{(M^2-m_V^{\ast 2})^2+m_V^{\ast 2}\Gamma_{\s{tot}}^{\ast 2}}
m_V^\ast\Gamma_{V\,\ra\,e^+\,e^-}^{\s{vac}}
\ea
where $\Gamma_{\s{tot}}^{\ast} = \Gamma_{V\,\ra\,{\s {all}}} 
- \Gamma_{{\s {all}}\,\ra V}$, $K_1$ is the modified Bessel function and
$\Gamma_{V\ra e^+e^-}^{\s{vac}}(M)$ is the partial width for the 
leptonic decay mode for off-shell vector particles. We use the above
equation to evaluate the invariant mass spectra of lepton pair originating 
from vector meson decays ($\rho\,\ra\,e^+\,e^-$ and $\omega\,\ra\,e^+\,e^-$).
The emission of dileptons from rho  and omega 
decay is affected due to changes in its mass and width 
(including collisional broadening due to $\omega\,\pi\,\ra\,\pi\,\pi$, 
$\rho\,\pi\,\ra\,\omega$ etc.)
at non-zero temperature and density.

Putting the different processes together it can be seen from the solid line 
in Fig.~\ref{totdilct} that the reduction in rho mass as compared to that of the
omega is well reflected in the dilepton spectrum even if one includes
the contribution from the pion annihilation channel $\pi\,\pi\,\ra
\,e^+e^-$. For comparison, the dashed line shows the sharp omega peak 
with $\omega\,\ra\,\rho\pi$ and $\omega\,\pi\,\ra\,\pi\,\pi$ closed. 
Also if one were to use unmodified
free meson properties it would be impossible to resolve the rho and omega 
peaks in the dilepton spectrum (dot-dashed curve).
\vskip 0.1in
\bef
\centerline{\psfig{figure=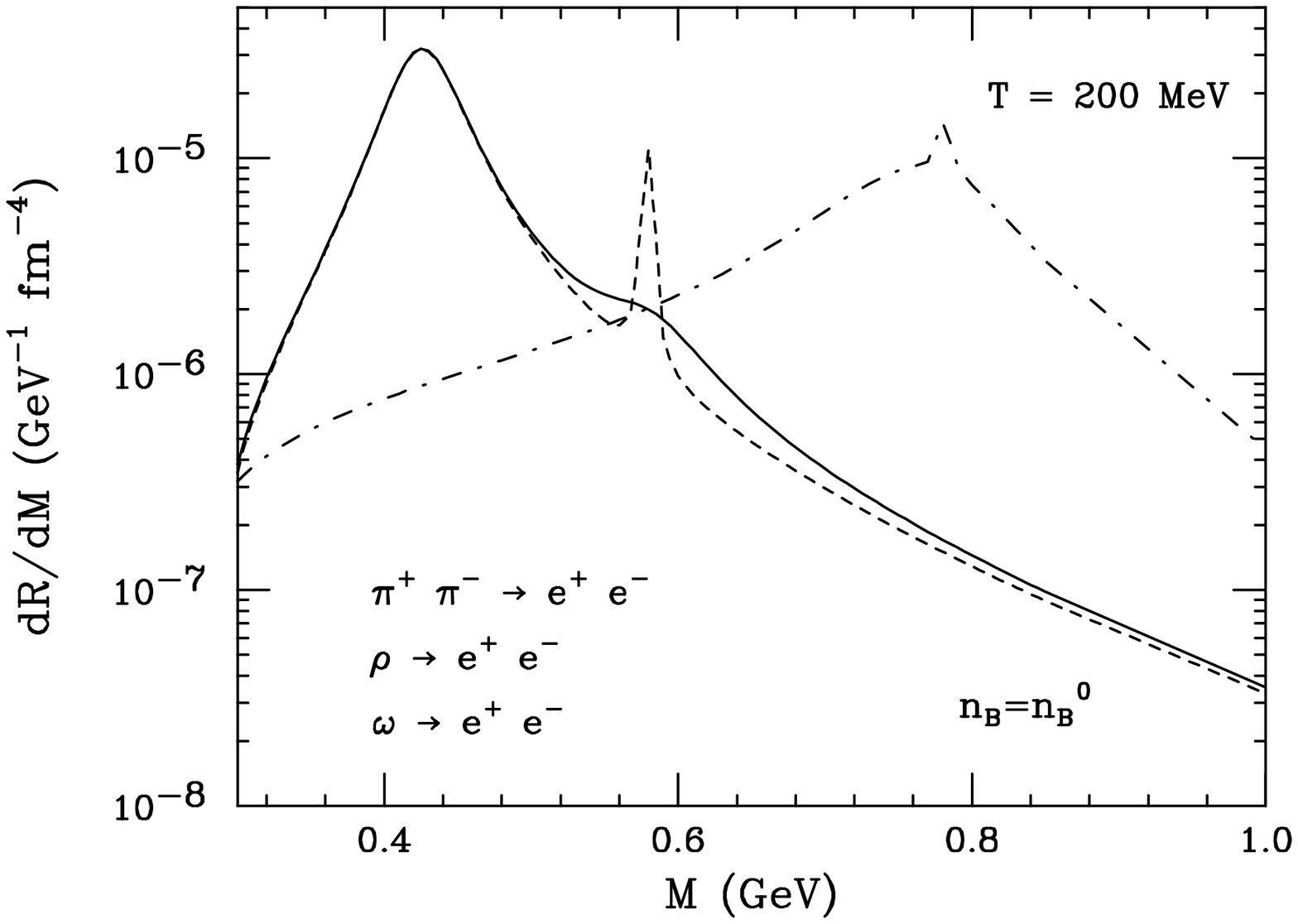,height=6cm,width=8cm}}
\caption{Invariant mass distribution of lepton pairs from
the decays $\omega\,\ra\,e^+\,e^-$,  $\rho\,\ra\,e^+\,e^-$ and the reaction
$\pi\,\pi\,\ra\,e^+\,e^-$ at $n_B=n_B^0$. The solid line corresponds to the 
case when $\Gamma_{tot}^\ast$ includes $\omega\,\ra\,3\pi$, 
$\omega\,\ra\,\rho\,\pi$ and $\omega\,\pi\,\ra\,\pi\,\pi$ for the
omega meson; $\rho\,\ra\,\pi\,\pi$ and $\rho\,\pi\,\ra\,\omega$ for
the rho meson. The dashed line shows the result with $\omega\ra\,3\pi$
only where the other two channels, as is often done, are ignored. 
Dot-dashed line indicates the yield without any medium effect whatsoever.
}
\label{totdilct}
\eef
\vskip 0.1in

The observed dilepton spectra originating from an expanding hadronic 
system is obtained by convoluting the 
static (fixed temperature) emission rate with the expansion dynamics.
In this work
we use the simple Bjorken-like model \cite{Bjorken} of boost invariant
longitudinal expansion to estimate the dilepton yield from an expanding
hadronic system.
In Fig.~\ref{totdil} we present the dilepton yield for initial
baryon density $n_B=2n_B^0$ (solid line) and
$n_B=n_B^0$ (dashed line) for $T_i=200$ MeV and $T_F=130$ MeV. 
The dot-dashed line indicates the dilepton spectrum when medium effects
on hadronic masses and decay widths are not considered.
The kink in the invariant mass plot of the dilepton 
yield at $M=680$ MeV (dashed curve), due to omega decay 
survives even after the space 
time evolution of the system is taken into account.  
Dramatic effects on the dilepton spectra due to the broadening of
omega should be discernible in experiments not involving expansion
scenario such as investigations with the upcoming HADES lepton pair
spectrometer at GSI.

The broadening of the $\omega$ meson due to different mechanisms
have been reported recently in the literature 
(see e.g.~\cite{Pisarski,Klingl,Wolf}). Pisarski~\cite{Pisarski} 
has argued that
the $\omega$-width could increase by an order of magnitude due 
to the reduction in pion decay constant connected with chiral
symmetry restoration, which in turn increases the coupling
$g_{\opr}$.
In another approach~\cite{Wolf} 
$\omega$ in motion with respect to the medium could couple 
through $N\bar{N}$ excitation with $\sigma$, which  subsequently decays to two
pion state, resulting a large broadening of $\omega$. 
However, if the $\omega$ is at rest such broadening of $\omega$ is absent, 
whereas the broadening mechanism proposed in the present work is possible 
even if the $\omega$ is at rest.
In the spectral function approach proposed by Klingl et al~\cite{Klingl}
$\omega$ becomes broad but still can be treated as a quasi particle,
supporting the basis of the present paper.
The opening of the new channel in a thermal bath has a contribution
similar to the channel $\omega\,\ra\,3\,\pi$, which is the most dominant
channel in vacuum, however, in a thermal bath the most dominant process
for the depletion of omega is $\omega\,\,\pi\,\leftrightarrow\,\pi\,\pi$.   
We emphasize at this point that for a short lived resonance ($R$)
which decays within the medium,
the width of the dilepton spectra is actually the rate at which 
it equilibrates 
($\Gamma_{\s{tot}}=\Gamma_{R\,\ra\,\s{all}} - \Gamma_{\s{all}\,\ra R}$),
involving in principle various processes in which $\s{R}$ participates.
It may be borne in mind that although elastic scattering contributes 
to kinetic equilibrium it has as such no direct effect on chemical 
equilibration of the system while of course such elastic processes are of
importance in phenomena such as viscosity etc. Indeed elastic scattering 
changes the momentum of the colliding particles but the nature of the
particles remains unaltered and hence this process does not contribute 
directly to the decay life time in the bath.
However, effects of elastic scattering on the broadening of vector
meson width has been considered in the literature~\cite{Haglin}.
 
Furthermore, the ratio
of the number of vector mesons decaying to lepton pairs inside 
($N_{\s{in}}$) to those decaying outside ($N_{\s{out}}$) the hot zone 
may be estimated to be (see e.g.~\cite{Hatsuda2}) 
$
N_{\s{in}}/N_{\s{out}} = (1-\exp(-\Gamma_{\s{tot}}R_T))/
\exp(-\Gamma_{\s{tot}}R_T)
$
where $R_T$ is the size of the system. For both $\rho$ and
$\omega$ (now with a considerably broadened omega) the above 
ratio turns out to be much greater than unity. This indicates that 
in the presence of nuclear matter at finite temperature a substantial
number of omega mesons decay inside the reaction volume and thus can
act both as a chronometer and a thermometer for hot hadronic matter
formed in relativistic heavy ion collisions.

Detailed measurement of the photoproduction of lepton pairs 
should provide invaluable insights into the creation,
propagation and decay of vector mesons inside the nuclear medium.
Changes in the rho and omega masses would reflect directly in the dilepton
invariant mass spectrum due to the quantum
interference between rho and omega mediated processes in the photoproduction
of lepton pairs (CEBAF). 
CERES collaboration ~\cite{Drees} 
has also planned to upgrade their experiment to improve the mass 
resolution so that the rho and omega may be disentangled under the 
condition mentioned above. 
It is argued~\cite{HBT} that the omega meson,
due to its long life time (23 fm/c), complicates the extraction 
of Hanbury-Brown Twiss (HBT) radii from the correlation function,
because the $\omega$ can distort the correlator to a highly 
non-Gaussian shape. An order of magnitude
reduction in the life time of the omega due to the $\rho\,\pi$ 
decay and $\omega\,\pi\,\ra\,\pi\,\pi$ reaction
in the medium should, however, remove such complications.
\vskip 0.1in
\bef
\centerline{\psfig{figure=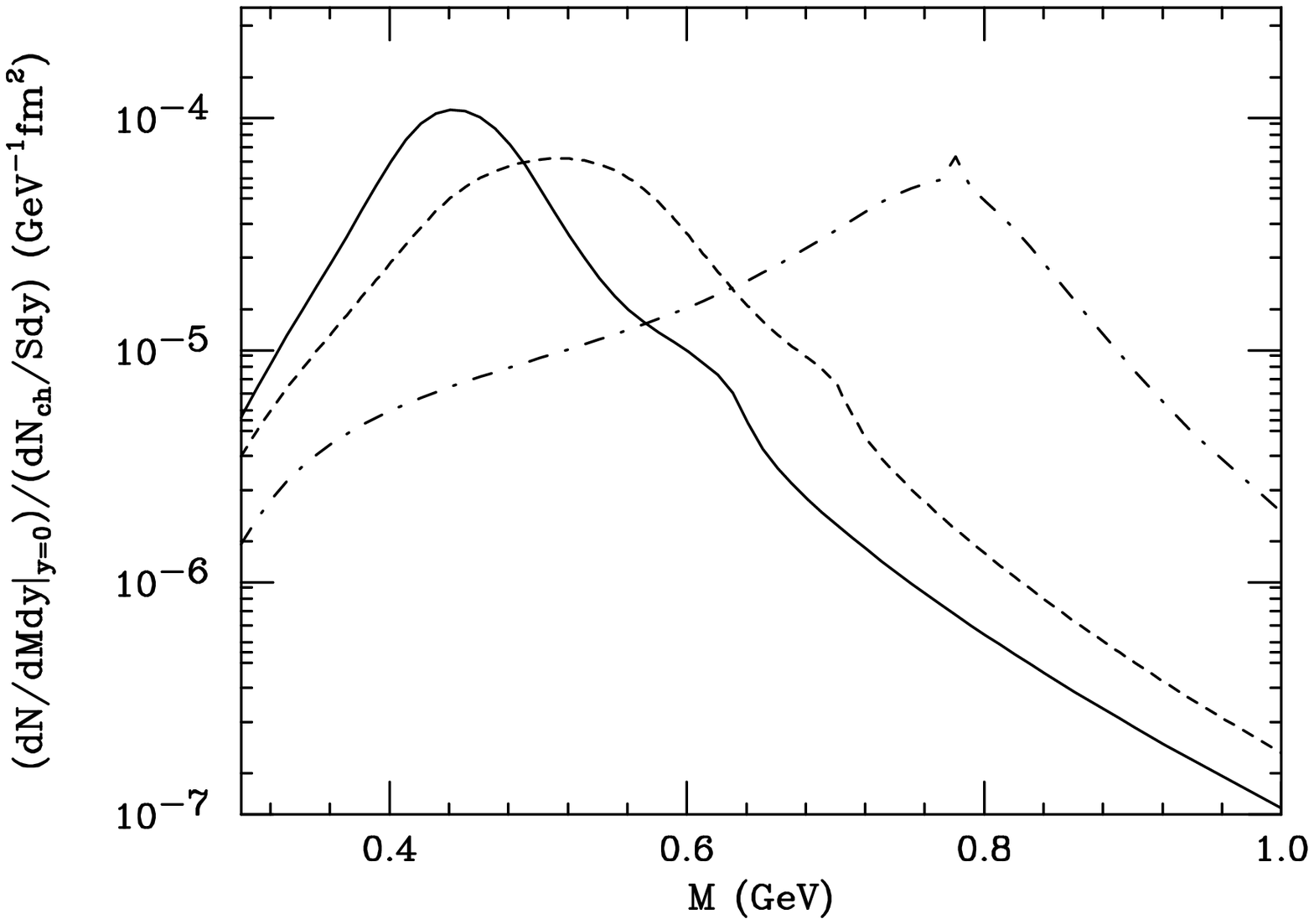,height=6cm,width=8cm}}
\caption{Invariant mass distribution for dilepton yield from
an expanding system for 
the decays $\omega\,\ra\,e^+\,e^-$, $\rho\,\ra\,e^+\,e^-$ and the reaction
$\pi\,\pi\,\ra\,e^+\,e^-$ (see text). $S$ and $dN_{ch}/dy$ denote 
the transverse area of the overlapped region and charge multiplicity
respectively. 
}
\label{totdil}
\eef

Therefore, our present observation has direct relevance to all those
experiments measuring dilepton emission from hot 
and dense matter in the low invariant mass region and 
HBT interferometry. A small mass difference between $\rho$ and $\omega$
mesons in vacuum makes it very difficult to disentangle the two peaks 
in the invariant mass spectra for dileptons. 
However, according to the present 
calculation, since the two peaks are widely separated,
it should be possible 
to observe  them through the radically altered shape of the
dilepton spectra. 
Various aspects of this novel phenomena and
its consequences on experimental observables are being pursued
vigorously.

\vskip .1in
\noindent{{\bf Acknowledgement:} We are grateful to Tetsuo Hatsuda
for useful discussions.}

\end{document}